\documentclass[10pt,twocolumn,letterpaper]{article}

\usepackage{3dv}
\usepackage{times}
\usepackage{epsfig}
\usepackage{graphicx}
\usepackage{amsmath}
\usepackage{amssymb}

\usepackage{graphicx}
\usepackage{color}
\usepackage{longtable}
\usepackage{booktabs}
\usepackage{bm}
\usepackage{amsmath}
\usepackage[ruled,vlined]{algorithm2e}
\usepackage{lipsum}
\usepackage{mathtools}
\usepackage{cuted}
\usepackage{flushend}
\usepackage{gensymb}
\usepackage{float}

\usepackage[pagebackref=true,breaklinks=true,letterpaper=true,colorlinks,bookmarks=false]{hyperref}

\threedvfinalcopy 

\def\threedvPaperID{93} 

\ifthreedvfinal\fi
\begin{document}

\title{V-NAS: Neural Architecture Search for Volumetric Medical Image Segmentation}

\author{Zhuotun Zhu\textsuperscript{1}, Chenxi Liu\textsuperscript{1}, Dong Yang\textsuperscript{2}, Alan Yuille\textsuperscript{1} and Daguang Xu\textsuperscript{2} \\
\textsuperscript{1}Johns Hopkins University, 
\textsuperscript{2}NVIDIA Corporation\\
{\tt\small \{zhuotun, alan.l.yuille\}@gmail.com}\quad
{\tt\small cxliu@jhu.edu}\quad
{\tt\small \{dongy, daguangx\}@nvidia.com}\quad \\
}

\maketitle

\begin{abstract}
Deep learning algorithms, in particular 2D and 3D fully convolutional neural networks (FCNs), have rapidly become the mainstream methodology for volumetric medical image segmentation. However, 2D convolutions cannot fully leverage the rich spatial information along the third axis, while 3D convolutions suffer from the demanding computation and high GPU memory consumption. In this paper, we propose to {\bf automatically} search the network architecture tailoring to volumetric medical image segmentation problem. Concretely, we formulate the structure learning as {\bf differentiable neural architecture search}, and let the network itself choose between 2D, 3D or Pseudo-3D (P3D) convolutions at each layer. We evaluate our method on $3$ public datasets, \emph{i.e.}, the NIH Pancreas dataset, the Lung and Pancreas dataset from the Medical Segmentation Decathlon (MSD) Challenge. Our method, named {\bf V-NAS}, consistently outperforms other state-of-the-arts on the segmentation tasks of both normal organ (NIH Pancreas) and abnormal organs (MSD Lung tumors and MSD Pancreas tumors), which shows the power of chosen architecture. Moreover, the searched architecture on one dataset can be well generalized to other datasets, which demonstrates the robustness and practical use of our proposed method.
\end{abstract}

\section{Introduction}


Over the past few decades, medical imaging techniques, \emph{e.g.}, magnetic resonance imaging (MRI), computed
tomography (CT), have been widely used to improve the state of preventative and precision medicine. With the emerging of deep learning, great advancement has been made for medical image analysis in various applications, {\em e.g.}, image classification, object detection, segmentation and other tasks. Among these tasks, organ segmentation is the most common area of applying deep learning to medical imaging~\cite{litjens2017survey}.



\begin{figure}[t]
    \centering
    \includegraphics[width=\linewidth]{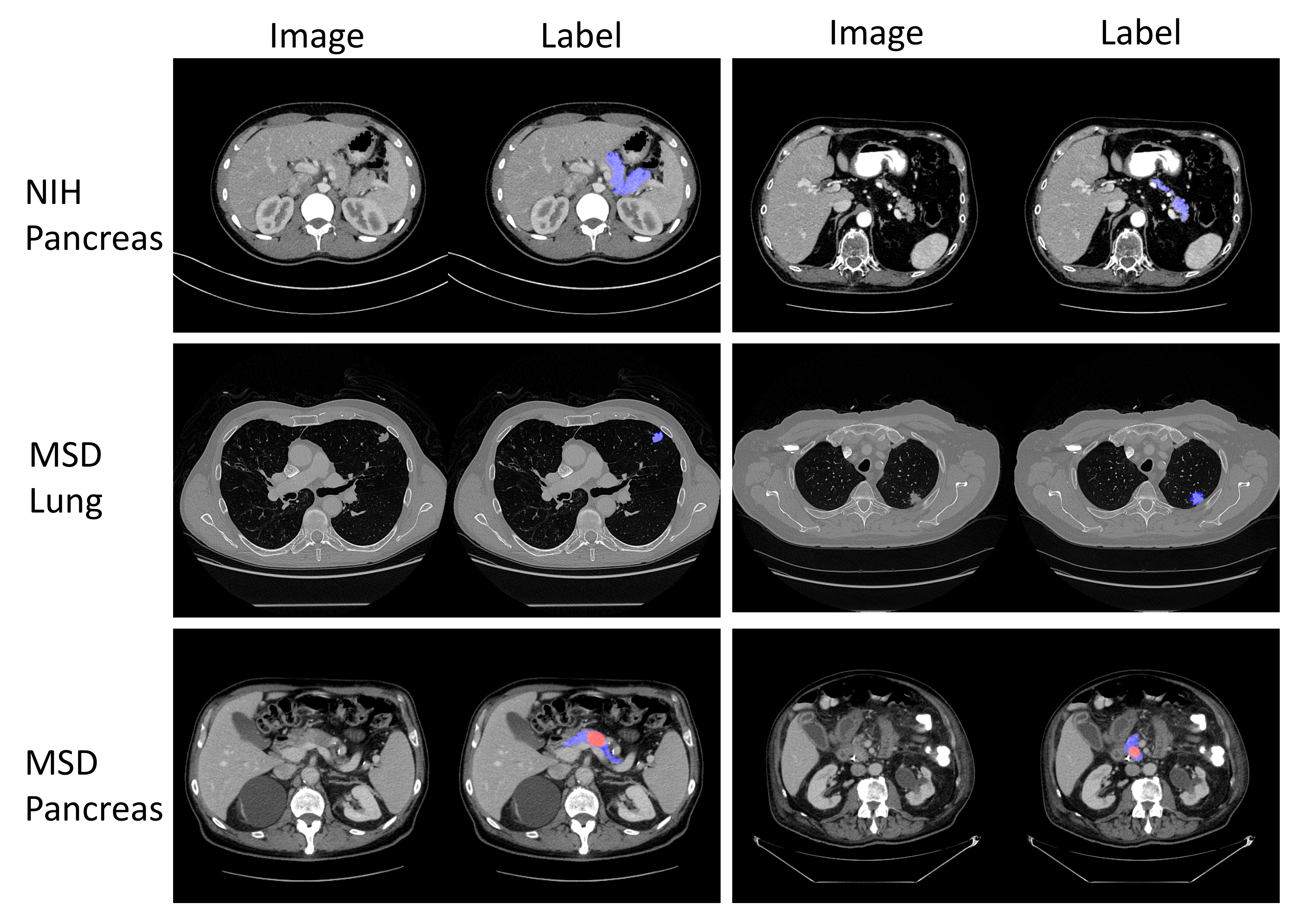}
    \caption{Typical examples from NIH Pancreas~\cite{roth2015deeporgan} in the $1$st row, MSD Lung Tumors~\cite{simpson2019large} in the $2$nd row and MSD Pancreas Tumors~\cite{simpson2019large} in the $3$rd row. Two slices of different cases are randomly chosen from each dataset. Normal Pancreas regions are masked as blue and abnormal pancreas regions are masked as red. The lung cancers are masked as blue. Best viewed in color.}
    \label{Fig:SegmentationTask}
\end{figure}

In this work, we focus on the volumetric medical image segmentation. Taking the pancreas and lung tumors segmentation from CT scans as an example as shown in Fig.~\ref{Fig:SegmentationTask}, the main challenges lie in several aspects: $1$) the small size of organs with respect to the whole volume; $2$) the large variations in location, shape and appearance across different cases; $3$) the abnormalities, \emph{i.e.}, the pancreas and lung tumors, can change the texture of surrounding tissues a lot; $4$) the anisotropic property along $z$-axis, which make the automatic segmentation even harder. 


To tackle these challenges, many segmentation methods have been proposed in the literature. Starting from handcrafted features, there are methods proposed to use intensity thresholding~\cite{akram2013multilayered}, region growing~\cite{pohle2001segmentation}, and deformable models~\cite{chandra2014focused}, which often suffer from the limited feature representation ability and are less invariant to the large organ variations. With a huge influx of deep learning related methods, fully convolutional neural networks (FCNs), \emph{e.g.}, 2D and 3D FCNs, have become the mainstream methodology in the segmentation area by delivering powerful representation ability and good invariant properties. The 2D FCNs based methods~\cite{cai2017improving,ronneberger2015u,roth2015deeporgan,roth2016spatial,zhou2017fixed} perform the segmentation slice-by-slice from different views, then fuse 2D segmentation output to obtain a 3D result, which is a remedy against the ignorance of the rich spatial information. To make full use of the 3D context, 3D FCNs based methods~\cite{cciccek20163d,dou20173d,milletari2016v,zhu2018a} directly perform the volumetric prediction. However, the demanding computation and high GPU consumption of 3D convolutions limit the depth of neural networks and input volume size, which impedes the massive application of 3D convolutions. Recently, the Pseudo-3D (P3D)~\cite{qiu2017learning} was introduced to replace 3D convolution $k$$\times$$k$$\times$$k$ with two convolutions, \emph{i.e.}, $k$$\times$$k$$\times$$1$ followed by $1$$\times$$1$$\times$$k$, which can reduce the number of parameters and show good learning ability in~\cite{liu20183d,wang2017automatic} on anisotropic medical images. However, all the aforementioned existing works choose the network structure empirically, which often impose explicit constraints, \emph{i.e.}, either 2D, 3D or P3D convolutions only, or 2D and 3D convolutions are separate from each other. These hand-designed segmentation networks with architecture constraints might not be the optimal solution considering either the ignorance of the rich spatial information for 2D or the demanding computations for 3D.


Drawing inspiration from recent success of Neural Architecture Search (NAS), we take one step further to let the segmentation network {\bf automatically} choose between 2D, 3D, or P3D convolutions at each layer by formulating the structure learning as {\bf differentiable neural architecture search}~\cite{liu2019auto,liu2018darts}. 
To the best of our knowledge, we are one of the first to explore the idea of NAS/AutoML in medical imaging field. Previous work \cite{mortazi2018automatically} used reinforcement learning and the search restricts to 2D based methods, whereas we use differentiable NAS and search between 2D, 3D and P3D, which is more effective and efficient. Without pretraining, our searched architecture, named V-NAS, outperforms other state-of-the-arts on segmentation of normal Pancreas, the abnormal Lung tumors and Pancreas tumors. In addition, the searched architecture on one dataset can be well generalized to others, which shows the robustness and potential clinical use of our approach.


\section{Related Work}
\subsection{\bf Medical Image Segmentation}\label{Sec:MedicalSegmentation}
The volumetric medical image segmentation has been dominated by deep convolutional neural networks based methods in recent years.~\cite{ronneberger2015u} proposed the UNet architecture tailored to tackle medical image analysis problems in 2D, which is based on an encoder-decoder framework: the encoder is designed to learn higher and higher level representations while the decoder decompresses compact features into finer and finer resolution to obtain dense prediction. Then, a similar approach was presented by~\cite{cciccek20163d} to extend UNet to 3D input. Later on, VNet~\cite{milletari2016v} proposed to incorporate residual blocks penalized by the Dice loss rather than the cross-entropy loss on 3D data, which directly minimizes the used segmentation error measurement. Meanwhile, a few recent works have been proposed to combine 2D and 3D FCNs as a compromise to leverage the advantages of both sides. \cite{xia2018bridging} adopted a 3D FCN by feeding the segmentation predictions of 2D FCNs as input together with 3D images. H-DenseUNet~\cite{li2018h} hybridized a 2D DenseUNet for extracting intra-slice features and a 3D counterpart for aggregating inter-slice contexts. Similarly, 2D FCNs and 3D FCNs are not optimized at the same time in~\cite{li2018h,xia2018bridging}. 


\subsection{\bf Neural Architecture Search}\label{Sec:NAS}

Neural Architecture Search (NAS) is the process of automatically discovering better neural architectures than human designs. 
We summarize the progress in along two dimensions: search algorithm and dataset/task. 

Many NAS algorithms belong to either reinforcement learning or evolutionary algorithm. 
In the reinforcement learning formulation~\cite{ZophVSL18}, the actions generated by an agent define the network architecture, and the reward is the accuracy on the validation set. 
In the evolutionary formulation~\cite{Real19}, architectures are mutated to produce better offsprings, again measured by validation accuracy. 
Although these algorithms are general, they are usually computationally costly. 
To address this problem,~\cite{LiuZNSHLFYHM18} progressively expand the search space in order to achieve better sample efficiency. 
Differentiable NAS approaches~\cite{liu2019auto,liu2018darts,ShinPS18} utilize sharing among candidate architectures, and are arguably the most efficient family of algorithms to date. 

At the same time, we also notice that the earlier papers~\cite{RealMSSSTLK17,XieY17,ZophL17} focused solely on MNIST or CIFAR10 dataset. 
Later,~\cite{LiuZNSHLFYHM18,Real19,ZophVSL18} searched for ``transferable architectures'' from the smaller CIFAR10 to the much larger ImageNet dataset. 
More recently,~\cite{Cai19,Zhang19} demonstrated the possibility to directly search for architectures on the ImageNet dataset. 
Finally,~\cite{liu2019auto} extended NAS beyond image classification to semantic segmentation. 

This paper sits at the frontier of both dimensions discussed above. 
We follow the differentiable NAS formulation for its efficiency. 
In terms of application domain, we directly search on volumetric image segmentation data, which is more demanding and challenging than 2D image labeling.

\section{Method}
\begin{figure}[t]
    \centering
    \includegraphics[width=\linewidth]{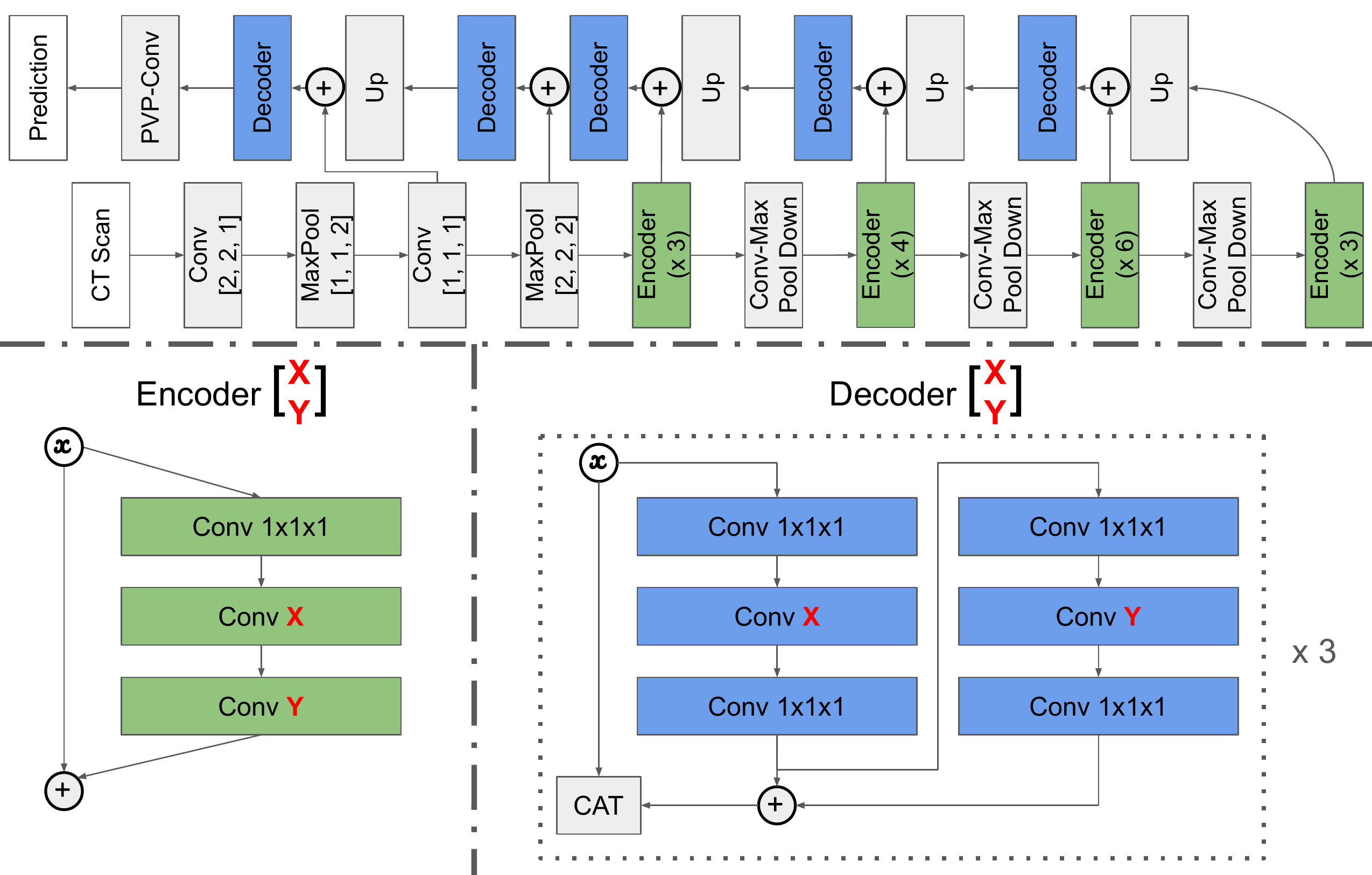}
    \caption{The segmentation network architecture. Each Encoder cell and Decoder cell has two candidate conv layers X and Y which are chosen between 2D, 3D, or P3D, whose details are defined in Sec.~\ref{Sec:EncoderSP} and Sec.~\ref{Sec:DecoderSP}. The Encoder along the encoding path is repeated by $3,4,6,3$ times while the decoder circled in the dashed rectangle is repeated by $3$ times. The encoder path is designed from ResNet-$50$, while the decoder path takes advantage of dense block and pyramid volumetric pooling (PVP). The first two convolutional layers adopt a kernel size $7$$\times$$7$$\times$$1$ with stride $[2,2,1]$ and $1$$\times$$1$$\times$$3$ with stride $[1,1,1]$. The overall network architecture is effectively verified by~\cite{liu20183d} while we add the searching process for color blocks to choose between 2D, 3D, and P3D.}
    \label{Fig:SegmentationNetwork}
\end{figure}

We define a {\bf cell} to be a fully convolutional module, typically composed of several convolutional (Conv+BN+ReLU) layers, which is then repeated multiple times to construct the entire neural network.
Our segmentation network follows the encoder-decoder~\cite{milletari2016v,ronneberger2015u} structure while the architecture for each cell, \emph{i.e.}, 2D, 3D, or P3D, is learned in a differentiable way~\cite{liu2019auto,liu2018darts}. The whole network structure is illustrated in Fig.~\ref{Fig:SegmentationNetwork}, where green Encoder and blue Decoder are in the search space.
We start with depicting the detailed network structure in Sec.~\ref{Sec:EncoderDecoderNetwork}, and then describing the search space of green Encoder and blue Decoder in Sec.~\ref{Sec:EncoderSP} and Sec.~\ref{Sec:DecoderSP}, respectively, followed by the optimization and search process in Sec.~\ref{Sec:Optimization}.

\subsection{Basic Network Architecture}\label{Sec:EncoderDecoderNetwork}
As shown in the upper part of Fig.~\ref{Fig:SegmentationNetwork}, our task is to train a convolution neural network model to predict the voxel labels of a CT scan as input. Similar to the state-of-the-art segmentation networks U-Net~\cite{ronneberger2015u}, V-Net~\cite{milletari2016v}, 3D U-Net~\cite{cciccek20163d} and ResDSN~\cite{zhu2018a}, our overall network structure consists of a high-to-low resolution process as a feature extractor, and then recovers the resolution through a low-to-high process to obtain dense predictions. To downsample 3D feature maps from a high resolution to a low resolution, the ``Conv-Max Pool Down" in the encoder path is implemented by a conv kernel of $1$$\times$$1$$\times$$1$ with a stride of $[2,2,1]$ followed by a MaxPool $1$$\times$$1$$\times$$2$ with a stride of $[1,1,2]$. The counterpart along the decoder path is realized by the ``Up" module to upsample 3D feature maps from a low resolution to a high resolution. More specifically, the ``Up" layer first projects the input feature map to match the number of feature channels of the higher Encoder
feature by a $1$$\times$$1$$\times$$1$ conv, followed by the 3D tri-linear interpolation and element-wise sum with the Encoder feature at a higher resolution. The residual connections from the lower-level encoder to the higher-level decoder aggregate more detailed information to semantic meaningful feature maps to give more accurate dense predictions. A pyramid volumetric pooling module~\cite{zhao2017pyramid} is stacked at the end of the decoder path before the final output layer for fusing multiscale features.

\subsection{Encoder Search Space}\label{Sec:EncoderSP}
The set of possible Encoder architecture is denoted as $\mathcal{E}$, which includes the following $3$ choices (\emph{c.f.}, Fig.\ref{Fig:SegmentationNetwork} for  $\text{Encoder}\begin{bmatrix}\textrm{X}\\ \textrm{Y}\end{bmatrix}$):
\begin{small}
\begin{equation}\label{Eq:EncoderCells}
    \{
    \underbrace{\text{Encoder}\begin{bmatrix}3\times3 \times 1 \\ 1 \times 1 \times 1\end{bmatrix}}_\text{$E_0$: 2D},
    \underbrace{\text{Encoder}\begin{bmatrix}3 \times 3 \times 3 \\ 1 \times 1 \times 1\end{bmatrix}}_\text{$E_1$: 3D},
    \underbrace{\text{Encoder}\begin{bmatrix}3 \times 3 \times 1 \\ 1 \times 1 \times 3\end{bmatrix}}_\text{$E_2$: P3D}
    \}
\end{equation}
\end{small}

%

As shown in Eq.~\ref{Eq:EncoderCells}, we define 3 Encoder cells, consisting of the 2D Encoder $E_0$, 3D Encoder $E_1$, and P3D Encoder $E_2$. $3$$\times$$3$$\times$$1$ is considered as 2D kernel. The input of the $l$-th cell is denoted as $x^l$ while the output as $x^{l+1}$, which is the input of the ${(l+1)}$-th cell. Conventionally, the encoder operation $O_e^l \in \mathcal{E}$ in the $l$-th cell is chosen from one of the $3$ cells, \emph{i.e.}, either $E_0$, $E_1$, or $E_2$. To make the search space continuous, we relax the categorical choice of a particular Encoder cell operation $O_e^l$ as a softmax over all $3$ Encoder convolution cells. By Eq.~\ref{Eq:RelaxationEncoder}, the relaxed weight choice is parameterized by the encoder architecture parameter $\bm{\alpha}$, where $\alpha_i^l$ determines the probability of encoder $E_i$ in the $l$-th cell,

\begin{equation}\label{Eq:RelaxationEncoder}
\begin{split}
    x^{l+1} &= O_e^l(x^l)\approx \bar{O}_e^l(x^l)\\ 
    \bar{O}_e^l(x^l) &= \sum\nolimits_{i=0}^2 \frac{\exp(\alpha_i^l)}{\sum\nolimits_{j=0}^2 \exp(\alpha_j^l)} E_i (x^l),
\end{split}
\end{equation}
where $l=1,\ldots, L$.

\subsection{Decoder Search Space}\label{Sec:DecoderSP}
Similarly, the set of possible Decoder architectures is denoted as $\mathcal{D}$, consisting of the following $3$ choices (\emph{c.f.}, Fig.~\ref{Fig:SegmentationNetwork} for $\text{Decoder}\begin{bmatrix}\textrm{X}\\ \textrm{Y}\end{bmatrix}$):
\begin{small}
\begin{equation}\label{Eq:DecoderCells}
    \{
    \underbrace{\text{Decoder}\begin{bmatrix}3 \times 3 \times 1 \\ 3 \times 3 \times 1\end{bmatrix}}_\text{$D_0$: 2D},
    \underbrace{\text{Decoder}\begin{bmatrix}3 \times 3 \times 3 \\ 3 \times 3 \times 3\end{bmatrix}}_\text{$D_1$: 3D},
    \underbrace{\text{Decoder}\begin{bmatrix}3 \times 3 \times 1 \\ 1 \times 1 \times 3\end{bmatrix}}_\text{$D_2$: P3D}
    \}
\end{equation}
\end{small}

As given in Eq.~\ref{Eq:DecoderCells}, we define $3$ Decoder cells, composed of the 2D Decoder $D_0$, 3D Decoder $D_1$, and P3D Decoder $D_2$. The Decoder cell is defined as dense blocks, which shows powerful representation ability in~\cite{li2018h,liu20183d}. The input of the $b$-th Decoder cell is denoted as $x^b$ while the output as $x^{b+1}$, which is the input of the ${(b+1)}$-th Decoder cell. The decoder operation $O_d^b$ of the $b$-th block is chosen from either $D_0$, $D_1$, or $D_2$. As shown in Eq.~\ref{Eq:RelaxationDecoder}, we also relax the categorical choice of a particular decoder operation $O_d^b$ as a softmax over all $3$ Decoder convolution cells, parameterized by the decoder architecture parameter $\bm{\beta}$, where $\beta_i^b$ is the choice probability of decoder $D_i$ in the $b$-th dense block,
\begin{equation}\label{Eq:RelaxationDecoder}
\begin{split}
    x^{b+1} &= O_d^b(x^b) \approx \bar{O}_d^b(x^b) \\
    \bar{O}_d^b(x^b)&= \sum\nolimits_{i=0}^2 \frac{\exp(\beta_i^b)}{\sum\nolimits_{j=0}^2 \exp(\beta_j^b)} D_i (x^b),
\end{split}
\end{equation}
where $b=1,\ldots, B$.

\subsection{Optimization}\label{Sec:Optimization}
After relaxation, our goal is to jointly learn the architecture parameters $\bm{\alpha}$, $\bm{\beta}$ and the network weights $w$ by the mixed operations. The introduced relaxations in Eq.~\ref{Eq:RelaxationEncoder} and Eq.~\ref{Eq:RelaxationDecoder} make it possible to design a differentiable learning process optimized by the first-order approximation as in~\cite{liu2018darts}. The algorithm for searching the network architecture parameters is given in Alg.~\ref{Alg:LearningProcess}. After obtaining optimal encoder and decoder operations $O_e^l$ and $O_d^b$ by discretizing the mixed relaxations $\bar{O}_e^l$ and $\bar{O}_d^b$ through $\texttt{argmax}$, we retrain the searched optimal network architectures on the $\mathcal{S}_{\text{trainval}} = \{\mathcal{S}_{\text{train}}, \mathcal{S}_{\text{val}}\}$ and then test it on $\mathcal{S}_{\text{test}}$.

\begin{algorithm}[t]
\DontPrintSemicolon
Partition the whole labeled dataset $\mathcal{S}$ into the {\bf disjoint} $\mathcal{S}_{\text{train}}$, $\mathcal{S}_{\text{val}}$ and $\mathcal{S}_{\text{test}}$;\;
Create the mixed operations $\bar{O}_e^l$ and $\bar{O}_d^b$ parametrized by $\alpha_i^l$ and $\beta_i^b$, respectively;\;
\While {training not converged}{
	1. Update weights $w$ by descending $\nabla_w \mathcal{L}_{\text{train}}(w, \bm{\alpha}, \bm{\beta})$\;
	2. Update $\bm{\alpha}$ and $\bm{\beta}$ by descending
	$\nabla_{\bm{\alpha}, \bm{\beta}} \mathcal{L}_{\text{val}}(w, \bm{\alpha}, \bm{\beta})$\;
}
Replace the relaxed operation $\bar{O}_e^l$ with $O_e^l = E_i, i = \texttt{argmax}_k {\exp(\alpha_k^l)}/{\sum\nolimits_{j=0}^2 \exp(\alpha_j^l)}$;\; 
Replace the relaxed operation $\bar{O}_d^b$ with $O_d^b = D_i, i = \texttt{argmax}_k {\exp(\beta_k^b)}/{\sum\nolimits_{j=0}^2 \exp(\beta_j^b)}$;\;
Retrain the discretized architecture on the $\mathcal{S}_{\text{trainval}}$.
\caption{V-NAS}
\label{Alg:LearningProcess}
\end{algorithm}

\section{Experiments}
\subsection{NAS Implementation Details}
In this work, we consider a network architecture with $L$=$3$+$4$+$6$+$3$=$16$ and $B$=$5$, shown as color blocks in Fig.~\ref{Fig:SegmentationNetwork}. The search space contains $3^{L+B}$=$3^{21}$$\approx$$10^{10}$ different architectures, which is huge and challenging. The architecture search optimization is conducted for a total of $40\rm{,}000$ iterations. When learning network weights $w$, we adopt the SGD optimizer with a base learning rate of $0.05$ with polynomial decay (the power is $0.9$), a $0.9$ momentum and weight decay of $0.0005$. When learning the architecture parameters $\bm{\alpha}$ and $\bm{\beta}$, we use Adam optimizer with a learning rate of $0.0003$ and weight decay $0.001$. Instead of optimizing $\bm{\alpha}$ and $\bm{\beta}$ from the beginning when weights $w$ are not well-trained, we start updating them after $20$ epochs. After the architecture search is done, we retrain weights $w$ of the optimal architecture from scratch for a total of $40\rm{,}000$ iterations. The searching process takes around $1.2$ V100 GPU days for one partition of train, val and test. All our models are trained on one V100 GPU with a customized batch size tuned to take full usage of the GPU memory due to different size input, which is computationally efficient in terms of neural architecture search task brought by the patch input.

In order to evaluate our method in the $4$-fold cross-validation manner to fairly compare with existing works, we randomly divide a dataset into $4$ folds, where each fold is evaluated once as the $\mathcal{S}_{\text{test}}$ while the remaining $3$ folds as the $\mathcal{S}_{\text{train}}$ and $\mathcal{S}_{\text{val}}$ with a train \emph{v.s.} val ratio as $2:1$. Therefore, there are in total $4$ architecture search processes considering the $4$ different $\{\mathcal{S}_{\text{train}}, \mathcal{S}_{\text{val}}\}$. The searched architecture might be different for each fold due to different 
$\{\mathcal{S}_{\text{train}}, \mathcal{S}_{\text{val}}\}$.
In this situation, the ultimate architecture is obtained by summing the choice probabilities ($\bm{\alpha}$ and $\bm{\beta}$) across the $4$ search processes and then discretize the aggregated probabilities. Finally, we retrain the optimal architecture on each $\mathcal{S}_{\text{trainval}}$ and evaluate on the corresponding $\mathcal{S}_{\text{test}}$.
All our implemented experiments use the same split of cross-validation and adopt Cross-Entropy loss, evaluated by the Dice-S{\o}rensen Coefficient (DSC) formulated as $\text{DSC}(\mathcal{P}, \mathcal{Y}) = \frac{2\times |\mathcal{P}\cap \mathcal{Y}|}{|\mathcal{P}| + |\mathcal{Y}|}$, where $\mathcal{P}$ and $\mathcal{Y}$ denote for the prediction and ground-truth voxels set for a foreground class, respectively. This evaluation measurement ranges in $[0, 1]$ where $1$ means a perfect prediction. We conduct experiments on $3$ public datasets, \emph{i.e.}, the NIH Pancreas dataset, the Pancreas and Lung dataset from the Medical Segmentation Decathlon (MSD) Challenge. And ablation studies are done on the NIH Pancreas dataset.


\begin{table*}[h]
\begin{center}
\begin{tabular}{lcccc}\toprule
Method           &{Categorization} & {Mean DSC}       		& {Max DSC}    		& {Min DSC} \\
\hline
{V-NAS (Ours)}  	&Search	& $\bm{85.15\pm{4.55}}\%$
& $91.18\%$       	&$\bm{70.37}\%$ \\
{Baseline (Ours)}  	&Mix	& $84.36\pm{5.25}\%$
& $91.29\%$       	&$67.20\%$ \\
\hline
{Xia \emph{et al.}~\cite{xia2018bridging}}  &2D/3D		& $84.63\pm{5.07}\%$ 				& $\bm{91.57\%}$       	&$61.58\%$ \\
{Zhu \emph{et al.}~\cite{zhu2018a}}  	&3D	& $84.59\pm{4.86}\%$ 				& $91.45\%$       	&$69.62\%$ \\
{Yu \emph{et al.}~\cite{yu2018recurrent}}  	&2D	& $84.50\pm{4.97}\%$ 				& $91.02\%$       	&$62.81\%$ \\
{Cai \emph{et al.}~\cite{cai2017improving}} &2D  & $82.40\pm{6.70}\%$          & $90.10\%$          & $60.00\%$          \\
{Zhou \emph{et al.}~\cite{zhou2017fixed}} &2D	& $82.37\pm{5.68}\%$ 				&$90.85\%$ 			& $62.43\%$\\
{Dou \emph{et al.}~\cite{dou20173d}} &3D &$82.25\pm{5.91}\%$ 				&$90.32\%$			&$62.53\%$ \\
{Roth \emph{et al.}~\cite{roth2016spatial}}	&2D	&$78.01\pm{8.20}\%$ 				&$88.65\%$			&$34.11\%$ \\
{Roth \emph{et al.}~\cite{roth2015deeporgan}} &2D		&$71.42\pm{10.11}\%$ 				&$86.29\%$			&$23.99\%$ \\
\bottomrule
\end{tabular}
\end{center}
\caption{
    Comparison with other state-of-the-arts on the NIH Pancreas dataset evaluated by the $4$-fold cross validation. Our one-stage segmentation network outperforms two-stage coarse-to-fine state-of-the-arts~\cite{xia2018bridging,zhu2018a}. The ``Categorization" column categorizes each method by whether the segmentation method is based on 2D, 3D, or by the dynamic searching in our proposed method. The architecture searched on the NIH Pancreas dataset is coded as [0 0 0, 0 0 0 1, 2 0 2 0 2 2, 0 0 0] for the 16 Encoder cells, and [0 0 1 0 1] for the 5 Decoder blocks.
}
\label{Tab:NIHPerformance}
\end{table*}

\subsection{NIH Pancreas Dataset}\label{Exp:NIH}
We conduct experiments on the NIH pancreas segmentation dataset~\cite{roth2015deeporgan}, which contains $82$ normal abdominal CT volumes. The size of CT volumes is $512\times512\times D$, where the number of slices $D$ is different for different cases, ranging in $[181, 466]$. The physical spatial resolution for one voxel is $w\times h\times d$, where $d = 1.0\textrm{mm}$ and $w = h$ that ranges from $0.5\textrm{mm}$ to $1.0\textrm{mm}$. For the data pre-processing, we simply truncate the raw Hounsfield Unit (HU) values to be in $[-100, 240]$ and then normalize each raw CT case to have zero mean and unit variance to decrease the data variance caused by the physical processes~\cite{gravel2004method} of medical images. As for the data augmentation in the training phase, we adopt simple yet effective augmentations on all training patches, \emph{i.e.}, rotation ($90\degree, 180\degree, \textrm{ and } 270\degree$) and flip in all three axes (axial, sagittal and coronal), to increase the number of 3D training examples which can alleviate the scarce of CT scans with expensive human annotations. Our training and testing procedure take patches as input to make more memory for the architecture design, where the training patch size is $96$$\times$$96$$\times$$64$ and the testing patch size is $64$$\times$$64$$\times$$64$ for the fine scale testing.

 
 

 
 As shown in Table~\ref{Tab:NIHPerformance}, our searched optimal architecture outperforms recent state-of-the-arts~\cite{xia2018bridging,yu2018recurrent,zhu2018a} segmentation algorithms.  It is well worth noting that state-of-the-arts~\cite{xia2018bridging,zhu2018a} adopt a two-stage coarse-to-fine framework to have an extra segmentation network to refine the initial segmentation maps whereas our method outperforms them by only one stage segmentation, which is more efficient and effective. We also obtain the smallest standard deviation and the highest Min DSC, which demonstrates the robustness of our segmentation across all CT cases. Furthermore, we implement the ``Mix" baseline that equally initializes all architecture parameters $\bm{\alpha}$ and $\bm{\beta}$ and keep them frozen during the training and testing procedures, which basically means the output takes exactly equal weight from 2D, 3D, and P3D in the encoder and decoder paths. Quantitatively, the search mechanism outperforms the ``Mix" baseline by $3.17\%$ and $0.79\%$ in terms of the Min and Mean DSC, respectively, which verifies the effectiveness of the searching framework. 
 
 In details, we code the searched optimal architecture on the NIH Pancreas dataset by [0 0 0, 0 0 0 1, 2 0 2 0 2 2, 0 0 0] for the 16 Encoder cells, and [0 0 1 0 1] for the 5 Decoder blocks, where ``0", ``1" and ``2" individually denote for the 2D, 3D and P3D, which are derived from definitions given in the Eq.~\ref{Eq:EncoderCells} and Eq.~\ref{Eq:DecoderCells}. We observe that 2D convolutions are mostly picked up in the beginning for encoders while P3D appears in the intermediate encoders, and 3D convolutions are mostly chosen in the ending decoders. We hypothesize that 2D layer is efficient to extract the within-slice information coupled with the P3D to fuse learned feature maps in the intermediate stage while 3D kernels are effective in the semantic meaningful layers close to the output prediction.

\begin{figure*}[h!]
	\begin{center}
		\includegraphics[width=0.9\linewidth]{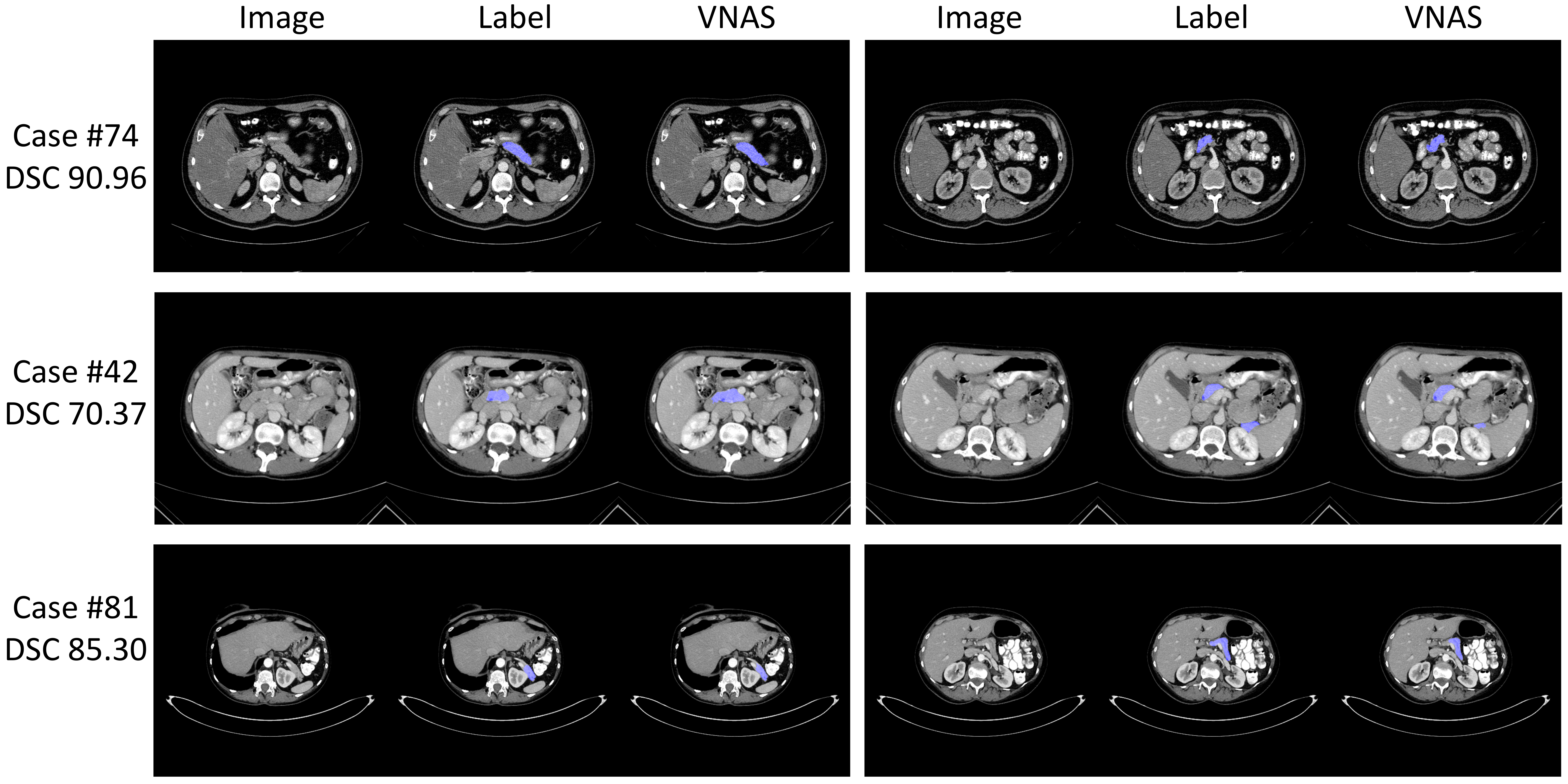} \\
	\end{center}
	\caption{The visualization illustration of predicted segmentation for ``VNAS" on the NIH Pancreas dataset. Two slices from Case ``\#74", ``\#42" and ``\#81" are randomly selected for visualization. The ``Min DSC" Case ``\#42" and an average DSC Case ``\#81" are chosen. Blue masked regions denote for the pancreas voxels. Best viewed in color.}
	\label{Fig:NIHPancreas}
\end{figure*}

 We visualize two slices randomly chosen from three NIH pancreas cases as shown in Fig.~\ref{Fig:NIHPancreas}. For the Case ``\#72" with a DSC of $90.96\%$, the pancreas appearance and boundary are well-captured and distinguished from its surroundings. For the Case ``\#81" with a DSC close to the ``Mean DSC", the pancreas regions are generally predicted well though with some minor under-estimations near the head. As for the Case ``\#42" with the min DSC, the ``VNAS" makes mistakes in the condition where the surrounding tissues are very complicate and the boundaries are ambiguous.

\subsection{MSD Lung Tumors}\label{Exp:MSDLung}
We also evaluate our framework on the Lung Tumors dataset from the Medical Segmentation Decathlon Challenge (MSD)~\cite{simpson2019large}, which contains $64$ training and $32$ testing CT scans. It is aimed for the segmentation of a small target (lung tumors) in a large image, where only the lung cancers are labelled and to be segmented. Since the testing label is not available and the challenge panel is currently closed, we report and compare results of $4$-fold cross-validation on the available $64$ training set. The truncation range is set to be $[-1000, 1000]$ to cover almost all the lung HU values in the data pre-processing while the data augmentation is the same as mentioned in Sec.~\ref{Exp:NIH}. More specifically, the patch size is set to be $64$$\times$$64$$\times$$64$ for the training and testing on MSD Lung Tumors dataset.


As given in Table~\ref{Tab:MSDLungPerformance}, our method (V-NAS-Lung) beats 3D UNet~\cite{cciccek20163d} and VNet~\cite{milletari2016v} by a large margin, at least $2.33\%$ in terms of the ``Mean DSC". The search process consistently outperforms the ``Mix" version which takes equally the 2D, 3D and P3D as a fixed configuration. It is worth noting that the ``Max DSC" of ours falls behind 3D UNet and VNet. We conjecture that since the overall network architecture is configured by the average choice probabilities of parameters $\bm{\alpha}$ and $\bm{\beta}$ on $4$ splits, our method tends to stably achieve the best overall segmentation performance, which is consistent with the much higher ``Median DSC". More specifically, the searched architecture on Lung tumors is coded as [0 0 0, 1 2 0 1, 2 1 2 0 0 0, 0 0 0] and [0 0 2 1 1].

To take one step further, we report results of directly training the searched optimal architecture from the NIH Pancreas dataset (V-NAS-NIH) on the MSD Lung tumors dataset from scratch. The searched architecture generalizes well and achieves better performance than ``Mix", 3D UNet and VNet. By comparing the two searched architectures from NIH Pancreas and MSD Lung Tumors datasets, we find that the two optimal architectures V-NAS-Lung and V-NAS-NIH share $68\%$ ($11$ out of $16$ Encoder cells) for the encoder path and $60\%$ ($3$ out of $5$ Decoder blocks) for the decoder path. The good property of transferring the network architecture searched on one dataset to another makes it possible for us to train the network architecture searched on a fairly big dataset with rich annotations to a small dataset with scarce annotations. We have not shown the ``Min DSC" in the table since all approaches miss some lung tumors considering the lowest DSC to be $0$, which shows that small lung tumors segmentation is a quite challenging task.



     \begin{table*}[h]
     \begin{center}
     \begin{tabular}{lcccc}\toprule
          Method &{Categorization} &{Mean DSC}& {Max DSC} & {Median}\\
    \hline
       V-NAS-Lung (Ours) &Search &$\bm {55.27\pm{31.18}}\%$ &$90.32\%$ &$66.95\%$\\
       V-NAS-NIH (Ours) &Search &$54.01\pm{31.39}\%$ &$92.17\%$ &$\bm{68.93\%}$\\
       Baseline (Ours) &Mix &$52.27\pm{31.40}\%$ &$89.57\%$ &$61.71\%$\\
    \hline
       3D UNet &3D &$52.94\pm{31.28}\%$ &$93.58\%$ &$61.08\%$\\
       VNet &3D &$50.47\pm{31.37}\%$ &$\bm{93.85\%}$ &$57.82\%$\\
     \bottomrule
     \end{tabular}
     \end{center}
     \caption{Performance of different methods on the MSD Lung tumors dataset evaluated by the same $4$-fold cross validation. The searched architecture on Lung tumors is coded as [0 0 0, 1 2 0 1, 2 1 2 0 0 0, 0 0 0] and [0 0 2 1 1]. It is worth noting that the searched architecture on the NIH dataset is well generalized to the Lung tumors dataset.}
     \label{Tab:MSDLungPerformance}
 \end{table*}

\subsection{MSD Pancreas Tumors}\label{Exp:MSDPancreas}
Different from the NIH normal pancreas dataset, the MSD Pancreas Tumors dataset is labeled with both pancreatic tumors and normal pancreas regions. The original training set contains 282 portal venous phase CT cases, which are randomly split into $4$ folds in our experiment, where each fold has its own training, validation and testing set and the final segmentation performance is reported on the average of $4$ folds. Since the resolution along $z$-axis of this dataset is very low and number of slices can be as small as $37$, the resolution of all cases on MSD Pancreas Tumors dataset are first re-sampled to an isotropic volume resolution of $d = 1.0\textrm{mm}$ for each axis. Then the pre-processing and data augmentation is the same as Sec.~\ref{Exp:NIH}. The patch size is set to be $64\times64\times64$ for both training and testing phases. Due to variant shapes and locations of tumors, the tumor segmentation is much more challenging and clinically important than the normal pancreas segmentation task since the early detection of pancreatic tumors can save lives.

As shown in Table~\ref{Tab:MSDPancreasPerformance}, our searched architecture consistently outperforms 3D UNet and VNet, especially the pancreas tumors DSC delivers an improvement of at least $1.79\%$, which is regarded as a fairly good advantage. The $7.68\%$ improvement over the manual ``Mix" setting on the pancreas tumors consistently proves the advantage of the architecture search in the volumetric image segmentation domain. In details, the searched architecture on this dataset is coded as [0 2 2, 2 0 0 0, 2 2 1 2 1 1, 0 1 1] and [1 0 2 0 1], by which we observe there are more P3D and 3D convolutions selected compared with the searched optimal architecture from the NIH normal Pancreas dataset. We hypothesize that the between-slice information is very important to detect abnormalities since radiologists need to scroll up and down when reading CT scans to help the diagnosis.

     \begin{table*}
     \begin{center}
     \begin{tabular}{lccccccc}\toprule
          Method &{Categor.} &\multicolumn{3}{c}{Pancreas Tumors DSC}&\multicolumn{3}{c}{Pancreas DSC}\\\cmidrule(lr){3-5}\cmidrule(lr){6-8}
           & &{Mean}& {Max} & {Median}&{Mean}& {Max} & {Min} \\
    \hline
       V-NAS (Ours) &Search &$\bm {37.78\pm{32.12}}\%$ &$92.49\%$ &$\bm{38.32\%}$&$\bm {79.94\pm{8.85}}\%$ &$\bm{92.24\%}$ &$36.99\%$\\
         Baseline (Ours) &Mix &$30.10\pm{31.40}\%$ &$92.95\%$ &$18.05\%$&$78.41\pm{9.40}\%$ &$92.21\%$ &$40.08\%$\\
        \hline
       3D UNet &3D &$35.61\pm{32.20}\%$ &$\bm{93.66\%}$ &$32.23\%$&$79.20\pm{9.43}\%$ &$91.95\%$ &$\bm{40.72\%}$\\
       VNet &3D &$35.99\pm{31.27}\%$ &$92.95\%$ &$35.91\%$&$79.01\pm{9.44}\%$ &$92.05\%$ &$28.15\%$\\
     \bottomrule
     \end{tabular}
     \end{center}
     \caption{Performance of different methods on the MSD Pancreas tumors dataset evaluated by the same $4$-fold cross validation. The results are given on the normal pancreas regions and pancreatic tumors, respectively. The searched architecture on Pancreas tumors dataset is coded as [0 2 2, 2 0 0 0, 2 2 1 2 1 1, 0 1 1] and [1 0 2 0 1].}
     \label{Tab:MSDPancreasPerformance}
 \end{table*}

\begin{figure*}[h!]
	\begin{center}
		\includegraphics[width=1.0\linewidth]{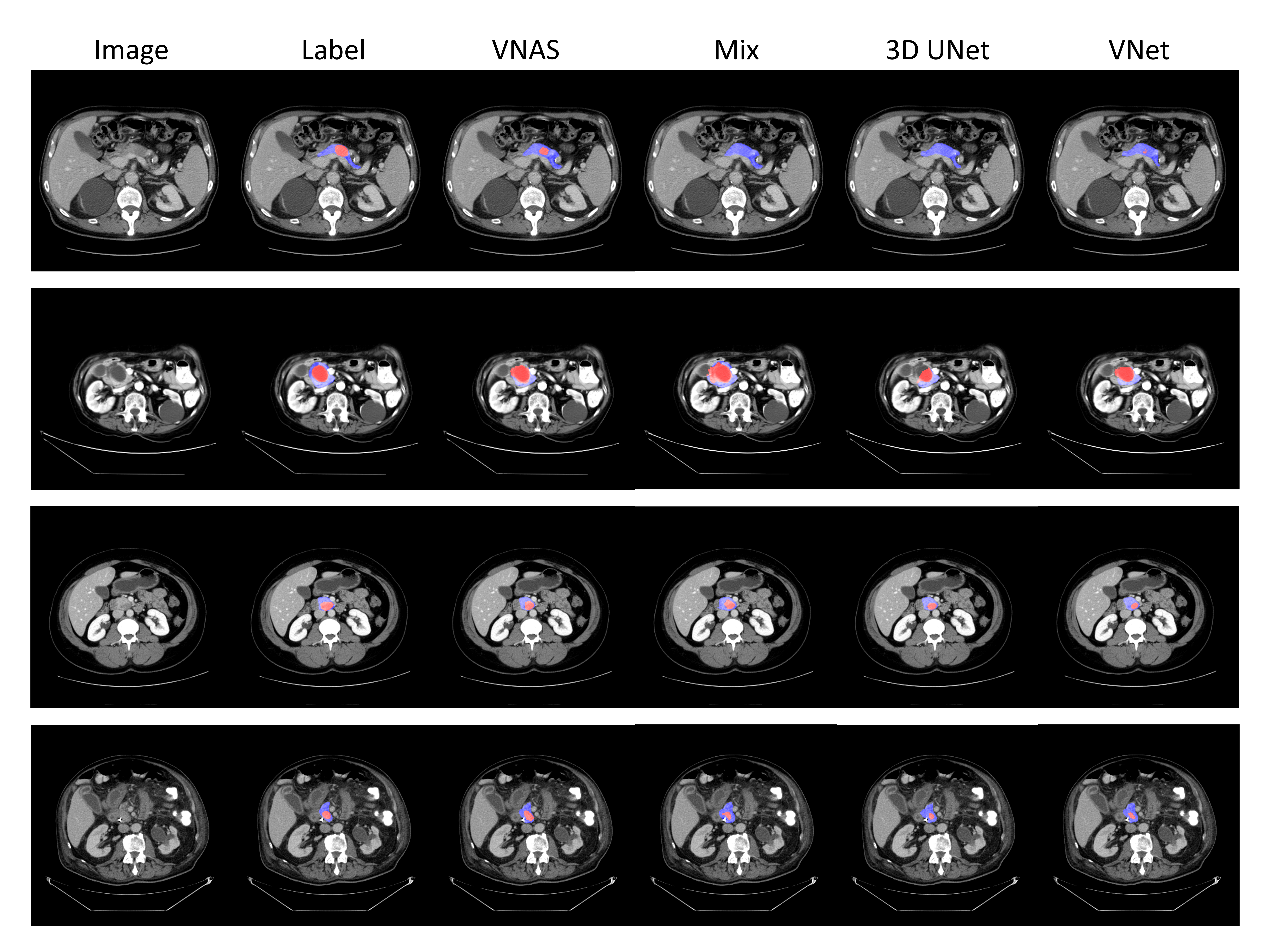} \\
	\end{center}
	\vspace{-0.5cm}
	\caption{The visualization illustration of predicted segmentation for ``VNAS", ``Mix", ``3D UNet" and ``VNet" on the MSD Pancreas Tumors dataset, which is the most challenging task among our $3$ segmentation tasks. Each row denotes a slice visualization from one case, and the specific cases numbers are ``309", ``021", ``069" and ``329" from top to bottom rows. The masked \textcolor{blue}{blue} and \textcolor{red}{red} regions denote for the normal pancreas regions and tumor regions, respectively. Best viewed in color.}
	\label{Fig:MSDPancreas}
\end{figure*}

We illustrate the visualization results of different methods as given in Fig.~\ref{Fig:MSDPancreas} on the same slice of a same case for comparison in each row. $4$ cases ( $\#309$, $\#021$, $\#069$ and $\#329$) are chosen from the MSD Pancreas dataset, which are shown from top to bottom at each row, respectively. Note that the masked red and blue regions denote the pancreas tumor and normal pancreas regions, respectively. For the case $\#309$ in the first row, the proposed ``V-NAS" successfully detects the tiny tumor regions while ``Mix" and ``3D UNet" totally fails and ``VNet" almost fails by only finding several tumor pixels. For the case $\#021$, $\#069$ and $\#329$ from the $2$nd to the $4$th row, the searched architecture can semantically capture the tumor regions better because it can adaptively leverage both the rich 3D spatial context, the 2D within-slice information and the anisotropic structures.

\section{Discussions}
To further verify the advantage of automatically selecting among 2D, 3D and P3D convolution layers via the neural architecture search, we conduct ablation studies on manually choosing types in encoder and decoder paths to be purely either 2D, 3D or P3D on NIH Pancreas and MSD Lung Tumors datasets in this section.
\subsection{Manual Setting on NIH Pancreas Dataset}\label{Dis:NIH}
  As shown in Table~\ref{Tab:NIHManualConfigurations}, we manually configurate the architecture of Encoder and Decoder, where we train and test all configurations on the same $4$-fold cross validation. More specifically, all Encoders are set to be one type (2D, 3D, or P3D), and the same strategy is applied to the Decoders.
 Each row denotes the pure categorical choice for the Encoder cells while the column for the Decoder. We can find that $2$D, $3$D, and P$3$D kernels contribute differently to the segmentation by the experimental results. The P3D as Encoder and the P3D as Decoder achieve a mean DSC of $84.75\%$ to outperform all other manual configurations. It is conjectured that the pure P3D takes most advantage of the anisotropic data annotation of the NIH dataset, where the annotation was done slice-by-slice along the $z$-axis. The different capability of learning semantic features between 2D, 3D and P3D for the dense volumetric image segmentation problem drives us to naturally formulate it to be a neural architecture search task. As it turns out, the automatic selection among the 2D, 3D and P3D delivers the best performance with a mean DSC of $85.15\%$ in Table~\ref{Tab:NIHPerformance}.
    \begin{table}[tb]
     \begin{center}
     \begin{tabular}{c|ccc}\toprule
          Encoder$\setminus$Decoder&3D  & 2D & P3D\\
    \hline
      3D &$84.09\%$ &$83.77\%$ &$84.20\%$\\
      2D &$83.66\%$ &$83.29\%$ &$84.08\%$\\
      P3D &$84.32\%$ &$84.69\%$ &\bm{$84.75\%$}\\
     \bottomrule
     \end{tabular}
     \end{center}
     \caption{Performance (``Mean DSC") of different encoder and decoder configurations on NIH dataset evaluated by the same $4$-fold cross validation. The architecture is manually set with different choices from 2D, 3D and P3D. Ours obtains $85.15\%$ in Table~\ref{Tab:NIHPerformance}.}
     \label{Tab:NIHManualConfigurations}
 \end{table}
 

     \begin{table}[h!]
     \begin{center}
     \begin{tabular}{lccc}\toprule
          Method &{Mean DSC}& {Max DSC} & {Median}\\
    \hline
       3D/3D &\bm{$53.74\pm{30.66}\%$} &$91.44\%$ &$60.55\%$\\
       2D/2D &$52.01\pm{31.50}\%$ &\bm{$92.58\%$} &$63.27\%$\\
       P3D/P3D &$51.48\pm{32.46}\%$ &$92.40\%$ &\bm{$63.89\%$}\\
     \bottomrule
     \end{tabular}
     \end{center}
     \caption{Performance of different encoder and decoder configurations on MSD Lung Tumors evaluated by the same $4$-fold cross validation. The architecture is manually configurated with different choices of 2D, 3D and P3D. Ours obtains $55.27\%$ in Table~\ref{Tab:MSDLungPerformance}.}
     \label{Tab:MSDLungManualPerformance}
 \end{table}


\subsection{Manual Setting on MSD Lung Tumors Dataset}\label{Dis:MSDLung}
On the MSD Lung Tumors dataset, we also report the manual architecture settings of 3D/3D, 2D/2D and P3D/P3D, \emph{e.g.}, ``3D/3D" stands for the configuration of only choosing 3D in both Encoder and Decoder cells. As given in Table~\ref{Tab:MSDLungManualPerformance}, the ``3D/3D" manual configuration achieves the best ``Mean DSC" of $53.74\pm{30.66}\%$. We suspect that the lung cancers are located inside the lung organs, which needs the rich spatial context to predict the abnormality. Consistent with what we observe in Sec.~\ref{Dis:NIH}, the neural architecture search idea outperforms all manual configurations to obtain a best mean DSC of $55.27\pm{31.18}\%$ with an advantage of $1.53\%$ over the ``3D/3D" in Table~\ref{Tab:MSDLungPerformance}.

\vspace{-0.25cm}
\section{Conclusion}
\vspace{-0.25cm}
We propose to integrate neural architecture search into volumetric segmentation networks to automatically find optimal network architectures between 2D, 3D, and Pseudo-3D convolutions. The search process is computationally efficient and effective. By searching in the relaxed continuous space, our method outperforms state-of-the-arts on both normal and abnormal organ segmentation tasks. Moreover, the searched architecture on one dataset can be well generalized to another one. In the future, we would like to expand the search space to hopefully find even better segmentation networks and reduce the computations.

{\bf Acknowledgements} We thank Huiyu Wang for his insightful discussions and suggestions.

\newpage
{\small
\bibliographystyle{ieee}
\bibliography{main}
}

\end{document}